\documentclass[12pt,aps,preprint,prb]{revtex4-1}
\usepackage{graphicx}
\usepackage{multirow}
\usepackage{color}

\begin{document}

\title{Quantum spin Hall states in graphene interacting with WS$_2$ or WSe$_2$}
\author{T. P. Kaloni$^1$}
\email{thaneshwor.kaloni@kaust.edu.sa, +1-2049522900}
\author{L. Kou$^2$}
\author{T. Frauenheim$^2$}
\author{U. Schwingenschl\"ogl$^{1,}$}
\email{udo.schwingenschlogl@kaust.edu.sa,+966-544700080}
\affiliation{$^1$Physical Science \& Engineering Division, KAUST, Thuwal 23955-6900, Kingdom of Saudi Arabia}
\affiliation{$^2$Bremen Center for computational Materials Science, University of Bremen, Am Falturm 1, 28359, Bremen, Germany}

\begin{abstract}
In the framework of first-principles calculations, we investigate the structural and
electronic properties of graphene in contact with as well as sandwiched between
WS$_2$ and WSe$_2$ monolayers. We report the modification of the band characteristics due to
the interaction at the interface and demonstrate that the presence of the dichalcogenides
results in quantum spin Hall states in the absence of a magnetic field.
\end{abstract}

\maketitle

\section{Introduction}

Research on graphene today appears to have reached a peak, mainly because the material
is rather difficult to be utilized in electronic devices due to limitations in
high quality mass production. Thus, interest is shifting to hybrid
systems with other two-dimensional materials, both for application purposes
and to create opportunities to better understand basic physical and chemical
phenomena \cite{geim,geim1}. Heterostructures of semiconducting MoS$_2$ and graphene already
have been demonstrated, showing potential particularly in data storage \cite{Bertolazzi}.
Moreover, the joint two-dimensional nature of the two components can be utilized for 
fabricating large scale flexible nanoelectronic devices. For example, a new generation
of field effect transistors based on heterostructures of WS$_2$ and graphene on transparent
and flexible substrates has shown a promising performance \cite{Georgiou}.

Two-dimensional topological insulators in a quantum spin Hall (QSH) state have
metallic edges that sustain dissipationless current flow \cite{Bernevig,Roth}. For graphene
this effect has been predicted by Kane and Mele using an analytical model \cite{kane}. 
Since the energy gap of graphene is tiny ($\sim10^{-3}$ meV) due to the small intrinsic
spin-orbit coupling (SOC) \cite{yao}, experimental observation of the QSH state is difficult.
Various ideas promising an enhancement have been put forward, all based on proximity to
areas in that the electrons are subject to strong SOC, including heavy
atom deposition \cite{prx}, H adsorption \cite{Balakrishnan}, and heterostructures
with MoTe$_2$ and Bi$_2$Te$_3$ \cite{arxiv}.

Experimental realization of a QSH state in the topological insulator HgTe
\cite{Bernevig} has initiated efforts on other materials, in particular on
graphene. Recently, experiments have demonstrated that graphene in a strong
magnetic field supports also a QSH state, which is
interesting for novel quantum circuits \cite{Young}. However, the experimental
difficulties coming along with large magnetic fields \cite{lau} call for alternative 
approaches. In this work we propose hybrid structures that host a QSH state in
the absence of a magnetic field. Specifically, we employ density functional theory
(including van der Waals corrections) to study the following systems: (a) graphene on
WS$_2$ or WSe$_2$ and (b) graphene sandwiched between two WS$_2$ or WSe$_2$ layers.
Our results prevail that the SOC in graphene can be enhanced such that a reasonable
energy gap is achieved at the K/K$'$ points with large band splitting. QSH states
are demonstrated for the sandwich systems.

\section{Computational details}

The heterostructures are constructed by joining a $4\times4\times1$ supercell of graphene
with $3\times3\times1$ supercells of the dichalcogenides. A vacuum layer of at least
15 \AA\ thickness guarantees that there is no artificial interaction perpendicular
to the two-dimensional system due to the periodic boundary conditions.
For the structural relaxation we employ the Quantum-ESPRESSO code \cite{qe},
using the generalized gradient approximation (GGA), in order to include the van der Waals forces.
The atomic positions are optimized until all forces have converged to less than 0.005 eV/\AA. 
The electronic structure calculations then are performed using the Vienna Ab-initio
Simulation Package \cite{vasp} both with pure GGA and including the spin. A plane wave 
cutoff energy of 450 eV and a Monkhorst-Pack $16\times16\times1$ k-mesh are employed.  

\section{Results and discussion}

\begin{figure}[ht]
\includegraphics[width=0.5\textwidth,clip]{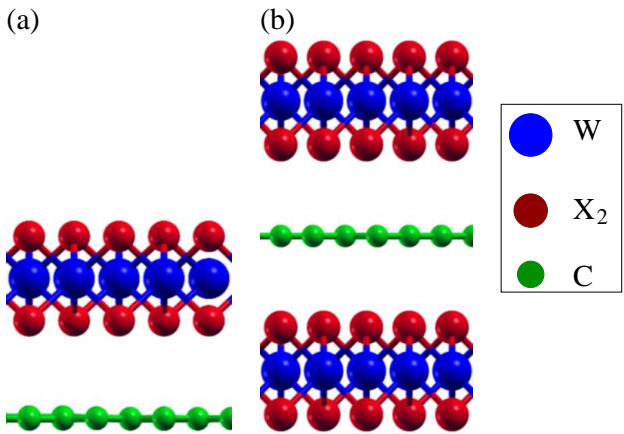}
\caption{(a) Graphene in contact with one dichalcogenide monolayer.
(b) Graphene sandwiched between two dichalcogenide monolayers.}
\end{figure}

\begin{table}[ht]
\caption{Distance between the subsystems ($d$), binding energy per C 
atom (E$_b$), band gap (E$_g$), valance band splitting ($\Delta_v$), and conduction band 
splitting ($\Delta_c$).} 
\begin{tabular}{|c|c|c|c|c|c|c|}
\hline
&\multicolumn{3}{|c|}{\multirow{1}{*}{GGA}}&\multicolumn{3}{|c|}{\multirow{1}{*}{GGA+SOC}}\\
\cline{2-5}
\hline
System&$d$ (\AA)& E$_b$ (meV)&E$_g$ (meV)& E$_g$ (meV)&$\Delta_v$ (meV)& $\Delta_c$ (meV)\\
\hline
Graphene/WS$_2$&3.41&53 &0.9&0.7&33&33 \\
\hline
WS$_2$/Graphene/WS$_2$&3.40&51&0.1&1.1&92&99 \\
\hline
Graphene/WSe$_2$&3.42&54&3.6&0.9&145 &132 \\
\hline
WSe$_2$/Graphene/WSe$_2$&3.42&52&3.0&1.0&149&153\\
\hline
\end{tabular}
\end{table}

The systems under consideration are illustrated in Fig.\ 1. The lattice parameters of
the graphene (9.98 \AA) and WS$_2$ supercells (9.57 \AA) give
rise to a lattice mismatch of 2.7\%. The distance between the two subsystems converges
to $d=3.41$ \AA, see Table I, which is a typical value for graphene on a semiconducting
substrate. We observe no modification of the C$-$C bond lengths, whereas the W$-$S bond lengths
(2.45 \AA\ to 2.46 \AA) are modified as compared to pristine WS$_2$
(2.39 \AA\ \cite{yun}) because of the lattice mismatch. We find $d=3.42$ \AA\ for
graphene on WSe$_2$ with a C$-$C bond length of 1.42 \AA, which is the value of pristine
graphene. The W$-$Se bond length (2.52 \AA\ to 2.53 \AA) is almost the same as in pristine
WSe$_2$ (2.52 \AA\ \cite{yun}). Moreover, we obtain $d=3.40$ \AA\ and $d=3.42$ \AA\
for graphene sandwiched between WS$_2$ and WSe$_2$ layers, respectively,
whereas the other structural parameters are the same as for the simple contacts.

Graphene is zero-gap semiconductor with very weak SOC \cite{Taroni}.
On the other hand, WS$_2$ is semiconductor with a direct band gap of 1.30-1.35 eV \cite{Braga}. 
Its structural and electronic properties have been studied extensively,
see Ref.\ \cite{Huang} and the references therein. WS$_2$ is widely utilized for
$n$/$p$-doped field effect transistors, for example \cite{Matthaus}.
Experimentally, the synthesis and characterization of graphene on WS$_2$ has been reported in
Ref.\ \cite{Georgiou}. The authors claim that a device based on this system
can operate at room temperature with good current modulating capacity.
We show in Fig.\ 2(a) the band structure obtained for graphene on WS$_2$ without SOC,
demonstrating some interaction between the two subsystems, though without chemical bonding.
The binding energy
\[E_{\rm binding}=E_{\rm graphene/WS_2}-E_{\rm WS_2}-E_{\rm graphene},\]
where $E_{\rm graphene/WS_2}$, $E_{\rm WS_2}$, and $E_{\rm graphene}$ are 
the total energies of the hybrid structure, WS$_2$, and graphene, respectively, 
amounts to 53 meV per C atom, in agreement with Ref.\ \cite{Zhang}. Due to this weak
interaction, a small band gap of 0.9 meV is obtained, see the zoomed view in Fig.\ 2(a),
with degeneracy of the K and K$'$ points.

When the SOC is switched on in the calculation, the spin degeneracy of the bands is
lifted, resulting in valence and conduction band splittings of 33 meV, see Table I,
with a small band gap of 0.7 meV, see the right hand side of Fig.\ 2(a). A similar
behavior has been observed for graphene on Bi$_2$Te$_3$ and MoTe$_2$ \cite{arxiv}.
The characteristic shape of the bands near the Fermi level signifies a QSH phase (band
inversion under preserved time-reversal symmetry), similar to the observations in
InAs/GaSb \cite{prl8}. It should be noted that the bands above and below the Fermi
level have opposite spin in agreement with the signature of the QSH phase. A similar
picture appears in germanene nanorods embedded in fully H-passivated germanene \cite{Seixas}
and in two-dimensional Sn \cite{Duan}, for which the gap can be strongly enhanced
by the application of the strain such that devices can operate at room temperature.

\begin{figure}[t]
\includegraphics[width=0.9\textwidth,clip]{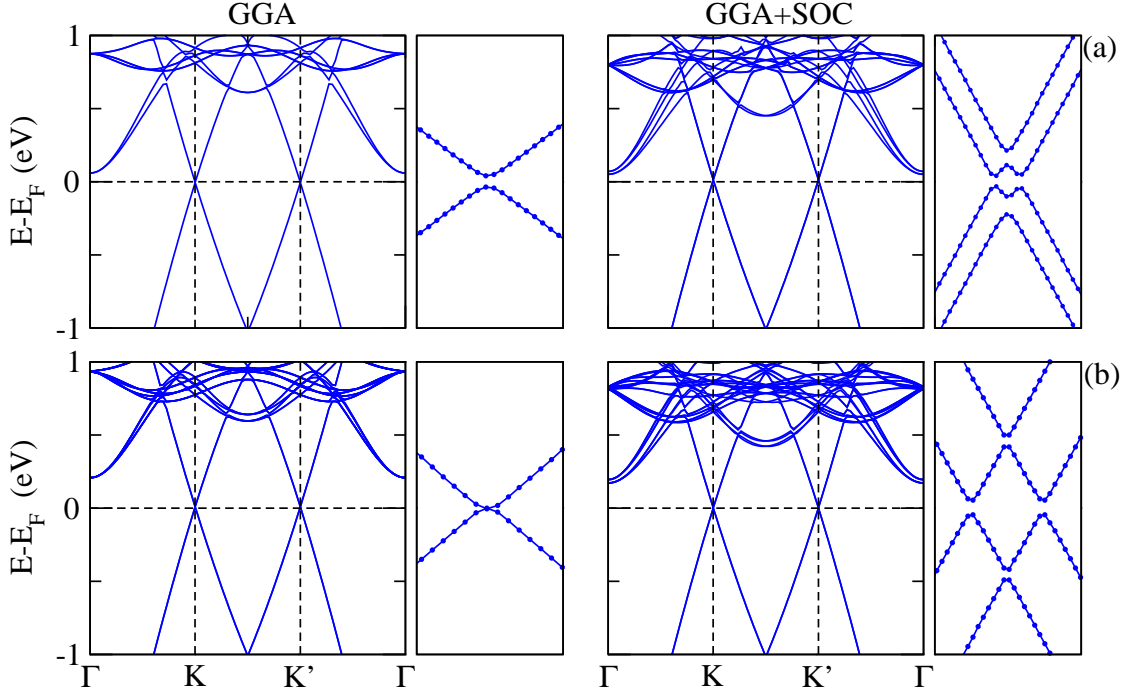}
\caption{GGA and GGA+SOC band structures for (a) graphene on WS$_2$ and (b)
graphene sandwiched between WS$_2$ layers. Zoomed views of the K point are shown
to clarify the band splitting.}
\end{figure}

In Fig.\ 2(b) we present results for graphene sandwiched between WS$_2$ layers. We find a 0.1
meV band gap without SOC and one of 1.1 meV when the SOC is taken into
account. The shape of the band structure again reflects a QSH state. Larger valence and
conduction band splittings of 92 meV and 99 meV, respectively, are achieved because two
WS$_2$ layers are attached to the graphene, see Table 1, similar to the 70 meV band
splitting of graphene in contact with the (111) surface of BiFe$_3$ (which is magnetic
and thus hosts a quantum anomalous Hall effect \cite{prl14}).

\begin{figure}[t]
\includegraphics[width=0.9\textwidth,clip]{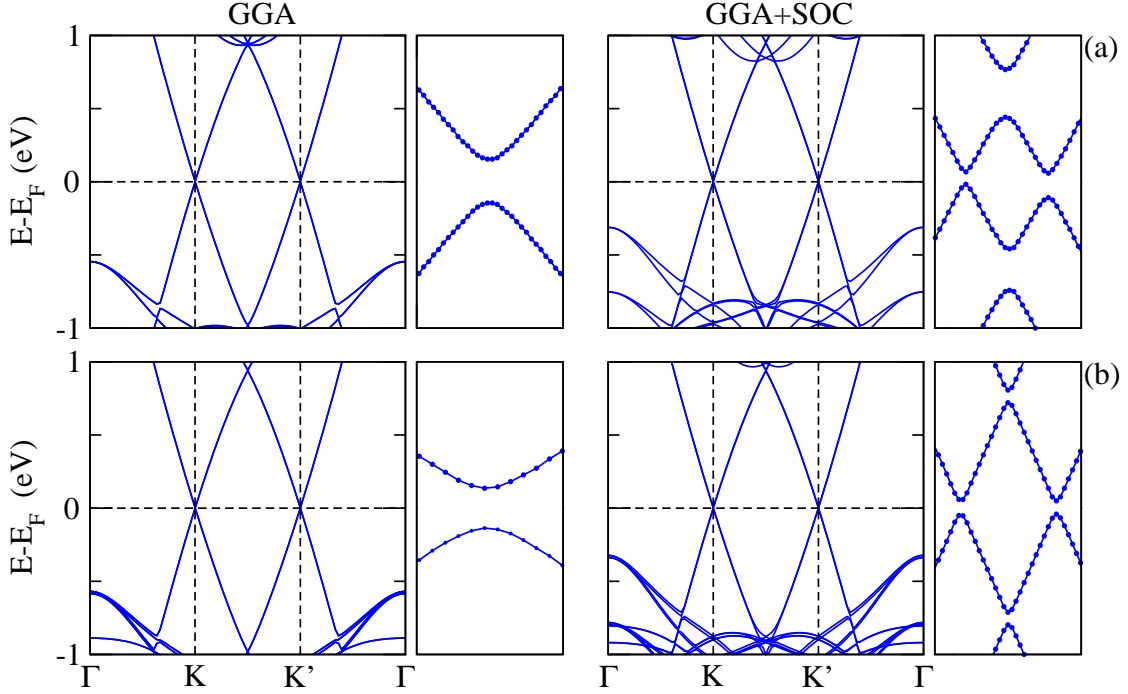}
\caption{GGA and GGA+SOC band structures for (a) graphene on WSe$_2$ and (b)
graphene sandwiched between WSe$_2$ layers. Zoomed views of the K point are shown
to clarify the band splitting.}
\end{figure}

We also study hybrid systems with monolayer WSe$_2$, which is a semiconductor with a
band gap of 1.65 eV \cite{chang}. The binding energy of graphene on WSe$_2$
turns out to be 54 meV and thus is similar to the WS$_2$ case. On the other hand,
the band gap of 3.6 meV, see Fig.\ 3(a), is about 5 times larger than for WS$_2$,
reflecting the stronger SOC in WSe$_2$. A comparison of the bulk properties
of WS$_2$ and WSe$_2$ can be found in Refs.\ \cite{amin,Rama}, for example.
Again, according to Fig.\ 3(a), the spin degeneracy at the K and K$'$ points is lifted
under inclusion of the SOC. The band structure qualitatively reflects the same
characteristics as demonstrated for WS$_2$ in Fig.\ 2(a). However, the valence and
conduction band splittings are enhanced to 145 meV and 132 meV, respectively, and
so is the band gap (0.9 meV), see the zoomed view in Fig.\ 3(a). This finding
is in agreement with the fact that the spin splitting due to SOC is about 50 meV
larger in bulk WeS$_2$ than in bulk WSe$_2$ \cite{amin,Rama}.
For graphene sandwiched between WSe$_2$ layers, we find a reduction of the band
gap from 3.0 meV to 1.0 meV under SOC with enhanced valence and conduction band splittings
of 149 meV and 153 meV, respectively, compare Fig.\ 3(b) to Fig.\ 2(b), as to be
expected from our previous discussion. In the WSe$_2$ systems the QSH effect therefore 
is significantly more pronounced than in the WS$_2$ systems.

\section{Conclusion}

Based on first-principles calculations, we have investigate the structural
and electronic properties of hybrid systems consisting of graphene and WS$2$ or WSe$_2$.
Band gaps of few meV are obtained due to the interaction between the component
materials. Moreover, band inversion is found at the K/K$'$ points, which indicates QSH
states. By the preserved time-reversal symmetry together with the enhancement of the
effective SOC in graphene, these systems are able to realize topological
phases and therefore QSH states. Usually a strong magnetic field 
is needed to achieve a QSH state in graphene \cite{Young}, while we
propose systems in that a QSH state appears without magnetic field, which
is a great advantage.

\end{document}